# Fizeau Drag in Graphene Plasmonics

Y. Dong[1,2]†, L. Xiong[1]†, I.Y. Phinney[3], Z. Sun[1], R. Jing[1], A.S. McLeod[1], S. Zhang[1], S. Liu[4], F. L. Ruta[1,2], H. Gao[3], Z. Dong[3], R. Pan[1], J. H. Edgar[4], P. Jarillo-Herrero[3], L.S. Levitov[3], A.J. Millis[1], M. M. Fogler[5], D.A. Bandurin[3]*, D.N. Basov[1]*

[1]Department of Physics, Columbia University, New York, NY, 10027, USA.
[2]Department of Applied Physics and Applied Mathematics, Columbia University
[3]Department of Physics, Massachusetts Institute of Technology, Boston, MA, 02139, USA.
[4]The Tim Taylor Department of Chemical Engineering, Kansas State University, Manhattan, KS, 66506, USA.
[5]Department of Physics, University of California San Diego, La Jolla, CA, 92093, USA.

† These authors contributed equally to this work.
* Corresponding authors:
D.A. Bandurin: bandurin.d@gmail.com
D.N. Basov: db3056@columbia.edu

**Dragging of light by moving dielectrics was predicted by Fresnel[1] and verified by Fizeau's celebrated experiments[2] with flowing water. This momentous discovery is among the experimental cornerstones of Einstein's special relativity and is well understood[3,4] in the context of relativistic kinematics. In contrast, experiments on dragging photons by an electron flow in solids are riddled with inconsistencies and so far eluded agreement with the theory[5,6,7]. Here we report on the electron flow dragging surface plasmon polaritons[8,9] (SPPs): hybrid quasiparticles of infrared photons and electrons in graphene. The drag is visualized directly through infrared nano-imaging of propagating plasmonic waves in the presence of a high-density current. The polaritons in graphene shorten their wavelength when launched against the drifting carriers. Unlike the Fizeau effect for light, the SPP drag by electrical currents defies the simple kinematics interpretation and is linked to the nonlinear electrodynamics of the Dirac electrons in graphene. The observed plasmonic Fizeau drag enables breaking of time-reversal symmetry and reciprocity[10] at infrared frequencies without resorting to magnetic fields[11,12], or chiral optical pumping[13,14].**

Graphene offers an ideal medium[15,16,17,18] for observing the plasmonic Fizeau drag as it supports the propagation of highly-confined, long-lived and electrically tunable surface plasmon polaritons. Crucially, graphene also withstands ultra-high current densities[19] so that the carrier drift velocity $u$ can be comparable with the SPP group velocity. In the absence of current, the SPP dispersion follows $\omega(-q) = \omega(q) \sim \sqrt{q}$ (Fig. 1a), where $\omega$ and $q$ are the frequency and wavevector of SPPs, respectively. Under applied current, the dispersion is predicted to depend [20,21,22,23,24] on the relative orientation of SPP propagation and carrier flow, generating the plasmonic Fizeau effect that breaks the reciprocity of the system: $\omega(-q) \neq \omega(q)$. Here we report on exploiting the unique attributes of graphene to demonstrate the physics of the plasmonic Fizeau effect where SPPs are dragged by drifting Dirac electrons.

In order to explore the plasmonic Fizeau drag, we fabricated multi-terminal graphene devices schematically shown in Fig. 1b. Monolayer graphene (MLG) encapsulated in hexagonal boron nitride (hBN) was integrated into back-gated structures assembled on a Si/SiO$_2$ substrate (285 nm of oxide). Gold SPP launchers[16,18] were deposited directly on graphene among the drain electrodes such that the current-gating effect [25] (Supplementary Notes "Transport and current-gating effects") from spatial inhomogeneity of electron density *n* was minimized. The gold launchers also served as ohmic contact to graphene and as heat sink for electrons in our high-current experiments. Finally, the width of the current-carrying graphene channel was narrowed down to 2 µm to boost the local current density and enhance the Fizeau drag effect. Real-space SPP images were acquired using low-temperature near-field optical nanoscopy techniques. An infrared (IR) laser of frequency ω illuminated the gold launcher (Fig. 1b) which launched propagating SPPs. The SPP electric field was out-coupled by a metallized tip of an atomic force microscope (AFM) into free-space photons and subsequently registered by a detector. Using the demodulated detector signal, the real-space profiles of SPPs were reconstructed[16,17,18]. The experiments were performed at cryogenic temperatures to reduce phonon-induced losses[18] and thus improve the fidelity of the measurements.

We now describe how the frequency-momentum dispersion ω(*q*) of SPPs is affected by the electric current in the graphene channel. The direct observable of our nano-imaging experiments is the wavelength λ$_p$ of SPPs, which is related to the real part of the SPP wavevector $q_1=2\pi/\lambda_p$ ($q= q_1+iq_2$). In realistic graphene devices, the square root law of SPP dispersion is modified by phonon resonances in the hBN substrate[16,17,18] but the dispersion relation ω(*q*) remains *q/-q* symmetric. Under applied direct current (dc) with density $J_{dc}$, the Dirac electrons supporting SPPs in graphene acquire a drift velocity *u*= $J_{dc}$/*en*, where *e* is the elementary charge and *n* is carrier density. The drifting carriers were predicted[20-24] to induce a plasmonic Fizeau effect which leads to an increase in λ$_p$ when SPP co-propagate with the carriers (Fig.1a, right branch) and decrease in λ$_p$ for counter-propagation scenario (Fig.1a, left branch). Our detailed theory (Supplementary Discussion) corroborates this intuition but reveals additional complications. Specifically, the drifting carriers modify the electromagnetic response of the system in a quasi-relativistic way[20,26] with the Lorentz factor $\gamma = (1 - u^2/v_F^2)^{-1/2}$ where $v_F$ is the graphene Fermi velocity. In addition, the corrections of Fizeau shift in the second order of the drift velocity are numerically significant prompting nonlinearities discussed below.

Next, we explore the experimental parameter space of the plasmonic Fizeau effect by modeling the wavelength shift Δλ$_p$/λ$_p$ for our device at different gate voltages $V_g$, laser frequencies *ω* and drift velocities *u* (Fig. 1c, d). Assuming a drift velocity of *u*=0.1$v_F$, a value readily achievable in our structures, the SPP wavelength is shortened compared to *u*=0 for plasmons counter-propagating with carriers (solid and dashed lines in Fig.1c). The magnitude of Fizeau shift Δλ$_p$/λ$_p$ increases at higher laser frequency *ω* and lower gate voltages $V_g$ (Fig. 1d). Choosing the lowest gate voltage $V_g$ = 30 V seems favorable for generating the largest Fizeau shift. However, both the amplitude and the propagation length of SPPs rapidly diminish at low $V_g$, compromising the fidelity of the experimental data. Higher laser frequency also results in similar adverse effects. We therefore chose to conduct the measurement at ω = 890 cm$^{-1}$ and a moderate gate voltage $V_g$= 47 V, corresponding to a carrier density $n$ = 2.9×10$^{12}$ cm$^{-2}$.

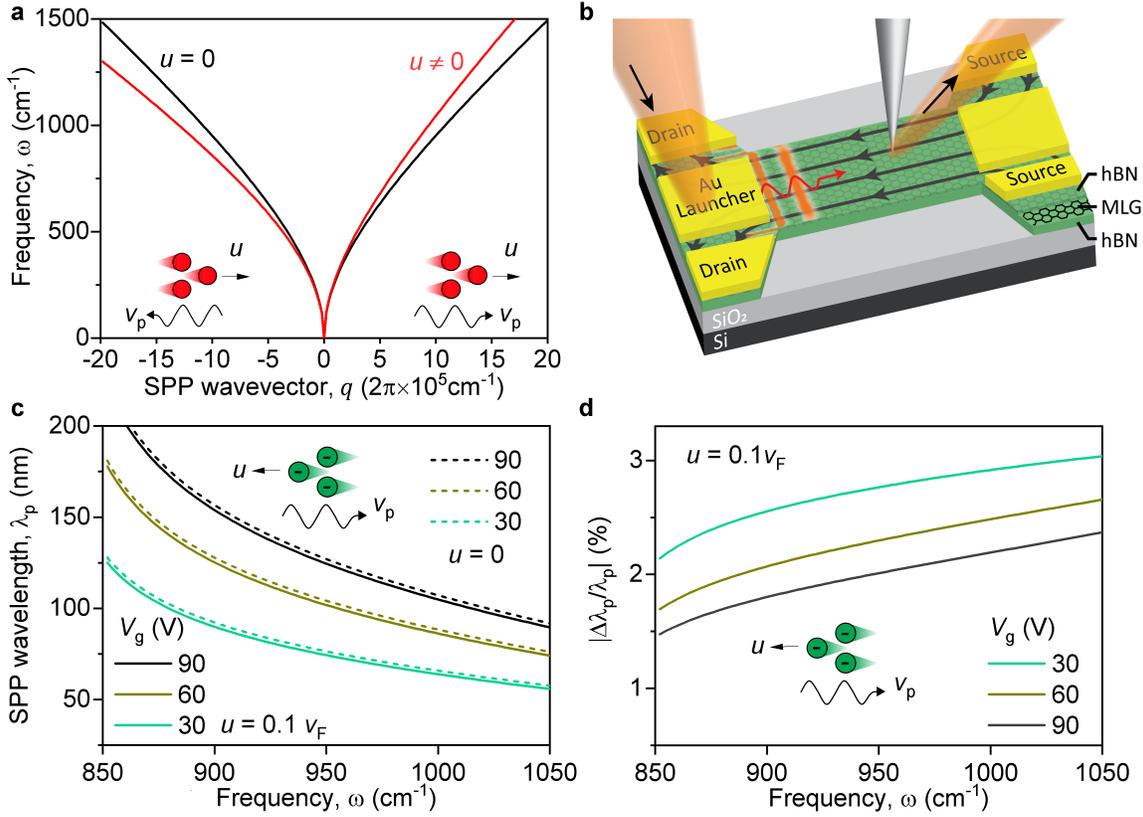

**Fig. 1. Plasmonic Fizeau drag in graphene: theory and modeling. a**, theoretically predicted SPP dispersion with (red line) and without (black line) drifting carriers. Carrier density is $n$ = 1×10$^{11}$ cm$^{-2}$ and the non-zero drift velocity is $u$=0.7$v_F$ with +$u$ along +$q$ direction. Wavy black arrows indicate SPP propagation direction and straight black arrows represents carrier drift direction. **b**, schematic of a graphene device with a constricted channel. Under the illumination of an IR laser, gold launcher excites propagating SPPs, which were visualized by near-field tip-based imaging. Black streamlines represent carrier drift directions. **c**, the SPP wavelength $\lambda_p$ as a function of the laser frequency ω with (solid lines) and without (dashed lines) current at different gate voltages $V_g$. The SPP wavelength is diminished under drifting current. **d**, the Fizeau shift $|\Delta\lambda_p/\lambda_p|$ as a function of the laser frequency ω at different gate voltages $V_g$ for a typical drift velocity $u = 0.1 v_F$.

We next analyze experimental evidence for the plasmonic Fizeau effect (Fig. 2) acquired using a representative device shown in Fig. 2a. Near-field signals were acquired at $T$ = 170 K and $V_g$ = 47 V by repeatedly scanning along the same line perpendicular to the launcher while varying the current density. Individual line scans were assembled into a two-dimensional false-color plot with position on the horizontal axis and current density on the vertical axis (Fig. 2b). In this representation, SPPs were launched by the gold launcher on the left of the field of view and propagated to the right, manifesting themselves as periodic oscillations[16,17,18] of the scattering amplitude signal (Fig. 2b).

Somewhat enhanced plasmonic loss can be observed at the largest applied current density, which is caused by Joule heating (Supplementary Notes "On possible extrinsic factors in the observed plasmonic wavelength shifts"). In Fig. 2c, we show line profiles averaged over a range of ±25µA/µm extracted from Fig. 2b at different current densities. Damped sinusoidal functions were used to fit (Supplementary Notes "Fitting method and statistical significance") the experimental data and the fitting results are displayed along with the raw data points (Fig. 2c). Experimental line profiles at different current densities show a smooth evolution of SPP wavelength with current density. For positive current densities, electrons flow towards the launcher while SPPs propagate away from the launcher. The counter-propagation of electrons and plasmons results in a clear reduction in the SPP wavelength, as evident from the comparison of the fitting results in Fig. 2d. The observed wavelength downshift is consistent with the $q<0$ dispersion branch in Fig. 1a where the SPP wavevector $q$ is enlarged in counter-propagation setting. Notably, the wavelength was not increased upon current polarity reversal due to second order correction to Fizeau effect described below. We acquired data from multiple devices at 170K and 60 K (Supplementary Data) which all showed the same trends as Fig. 2b-d. The fact that the SPP wavelength depends on current polarity suggests that the SPP Fizeau effect breaks the $q/-q$ reciprocity in our nano-plasmonic device.

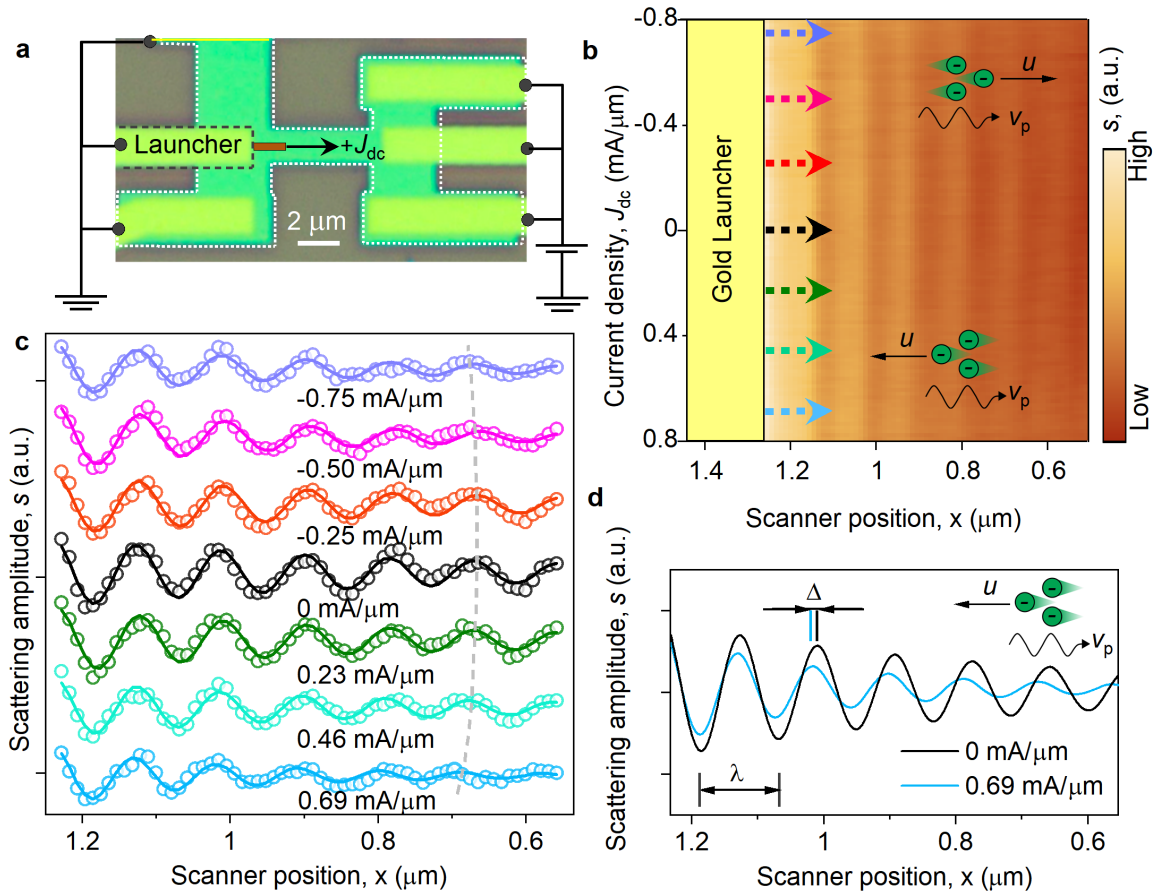

**Fig. 2. Experimental demonstration of the plasmonic Fizeau drag. a**, an optical image of a representative device. Data displayed in panels **b-d** were collected in the dark red region

while scanning along a single line. The black arrow represents positive current direction. **b**, Near-field image, at $V_g$ = 47 V and $T$ = 170 K, acquired by scanning along the same line in panel **a** while varying the current density between ±0.8mA/µm. The dashed arrows represent positions where the averaged line profiles in panels **c** and **d** were taken. One-dimensional Fourier filter was applied for b only to reduce visual noises. **c**, Averaged (±25µA/µm) SPP line profiles at different current densities. The circles are raw data; the solid lines are fitted results; the dashed line is a guide to the eye. The line profiles are shifted vertically for clarity. **d**, Fitted line profiles of the SPP without dc current (black) and with $J_{dc}$=0.69 mA/µm (blue) illustrating the reduction of the SPP wavelength.

Having established qualitative indicators of the plasmonic Fizeau drag, we now examine the findings quantitatively. We fit every line profile (Supplementary Notes "Fitting method and statistical significance") in data sets such as Fig. 2b and obtain the Fizeau shift $\Delta\lambda_p/\lambda_p$ as a function of current density $J_{dc}$ and drift velocity $u$ (Fig. 3a and b). The results reveal that the Fizeau shift reaches -2.5% at $J_{dc}$ = 0.7 mA/µm for counter-propagation setting (Fig. 3a). For co-propagation setting, the Fizeau shift is vanishingly small within the experimental error. When reversing the polarity of the gate voltage, which switches the carrier type to holes, the observed dependence of $\Delta\lambda_p/\lambda_p$ on hole drift velocity was similar to that for electrons (Fig. 3b). These observations indicate an ambipolar character of the Fizeau drag, which depends on the carrier drift velocity but not on the carrier type.

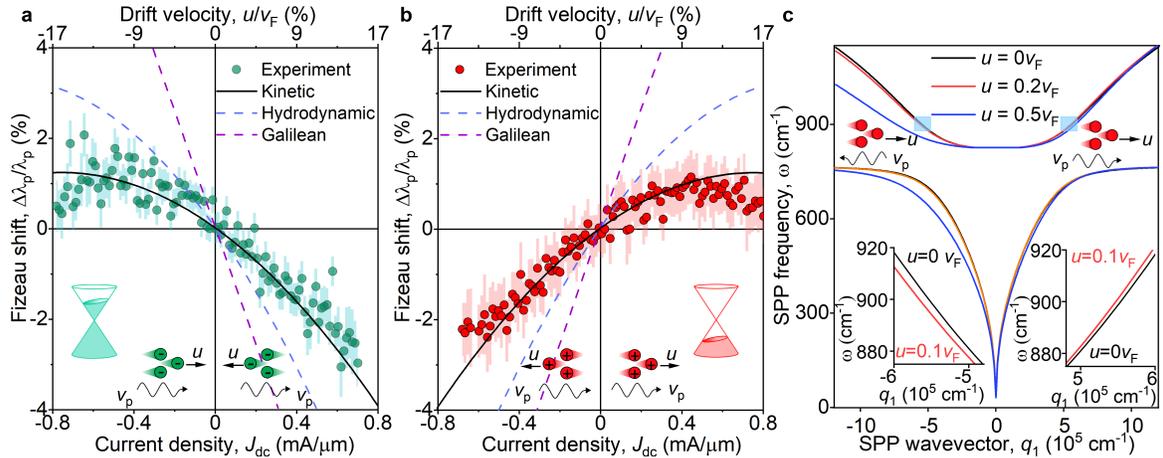

**Fig. 3. Quantitative analysis of the plasmonic Fizeau drag. a**, the Fizeau shift (circles) $\Delta\lambda_p/\lambda_p$ extracted from fitted line profiles as a function of current density and carrier drift velocity at $V_g$ = 47 V and $T$=170K. Error bars represent ±1 standard deviation of the fitted Fizeau shift. Lines represent theoretical predictions of Fizeau shift for the kinetic regime (black solid line), the hydrodynamic regime (blue dashed line), and a 2D electron gas with parabolic dispersion (purple dashed line), respectively. **b,** same plot as panel **a** at $V_g$ = -47 V and $T$= 170K. **c,** SPP dispersion with carrier density $n$ = 2.9×10$^{12}$ cm$^{-2}$ calculated for the kinetic regime (Supplementary Discussion). The gap in the dispersion stems from phonon resonances in the hBN. Insets show enlarged view of the regions marked by the light blue

boxes.

A salient feature of the experimental drag data is that the shifts in Fig. 3a, b are neither symmetric with the current direction nor linearly varying with the carrier drift velocity. The observed experimental trends can be understood based on a model which accounts for both linear and quadratic terms of the Fizeau shift as a function of $u \ll v_F$:

$$\frac{\Delta\lambda_p}{\lambda_p} = \eta\frac{u}{v_g} - \left(\eta + \frac{1}{4}\right)\frac{v_p}{2v_g}\frac{u^2}{v_F^2}, \quad v_p = \frac{\omega\lambda_p}{2\pi} \quad (1)$$

Note that $v_g \approx \frac{1}{2}v_p$ reduces to the plasmon group velocity only when the dielectric screening is not dispersive. In our device, the screening from hBN boosts the magnitude of Fizeau shift by suppressing $v_g$. The drag coefficient $\eta$ assumes a value in the interval $1/4 \leq \eta \leq 1/2$ depending on the quasiparticle collision rate $\Gamma_{ee}$ (Supplementary Discussion). In particular, $\eta = 1/4$ is expected provided $\Gamma_{ee}$ is small compared to the measurement frequency $\omega = 890$ cm$^{-1} = 26.7$ THz. The prediction of Eq. (1) for this kinetic regime is plotted by the black solid lines in Fig. 3a, b. Provided $\Gamma_{ee} > \omega$, the regime where graphene quasiparticles behave collectively as a hydrodynamic fluid[26, 27,31], the Fizeau coefficient assumes $\eta = 1/2$, which is plotted by blue dashed lines in Fig. 3a, b. Here we ignore, for simplicity, the Fermi-liquid interaction effects[28]. For our experimental condition, we estimate that $\Gamma_{ee} \sim T^2/|\hbar v_F \sqrt{n}| < 1$ THz, which strongly suggests that $\eta = 1/4$ should be appropriate, in agreement with the experimental data. It is also instructive to compare these results with previous works (Supplementary Table) on Fizeau shift (often referred to as the Doppler shift) of plasmons in plasma[5, 29] and in GaAs semiconductor structures[6,7, 30]. Such systems have parabolic quasiparticle dispersions, which are invariant under a Galilean transformation. However, graphene quasiparticles are massless Dirac fermions characterized by a quasi-Lorentz invariance, with the speed of light replaced by $v_F$. As indicated by the purple dashed lines in Figs. 3a, b, the Fizeau shift of plasmons in a Galilean system should be linear with the drift velocity, with the unit drag coefficient $\eta = 1$. The obvious discrepancy of these predictions with our experimental data vividly demonstrates that the Galilean invariance is broken in graphene.

Additional insights into the Fizeau shift can be obtained by placing our results in the context of the nonlinear electrodynamics of graphene[31]. The current response at the plasmonic frequency $\omega$ can be expanded as $j(\omega) = \left(\sigma + \sigma^{(2)}E_{dc} + \sigma^{(3)}E_{dc}^2 + \cdots\right)E_{ac}(\omega) \equiv \sigma_{eff}E_{ac}(\omega)$ where $\sigma^{(n)}$ is the $n$-order nonlinear optical conductivity component, $E_{dc}$ is the dc electric field that drives the static current flow $u \sim E_{dc} + O(E_{dc}^3)$ and $E_{ac}(\omega)$ is the inhomogeneous ac field from the launcher. All orders of $E_{dc}$ contribute to the effective ac conductivity $\sigma_{eff}$ which determines the Fizeau shift. Moreover, the Fizeau shift caused by $E_{dc}^k$ is a signature of $k+1$-order nonlinearity represented by the component $\sigma^{(k+1)}$. Our data in Fig. 3 show that as the dc drive increases, the linear Fizeau shift acquires quadratic corrections: a manifestation of the 3$^{rd}$-order nonlinearity in $\sigma^{(3)}E_{dc}^2$.

The current-induced Fizeau drag of plasmons reveals novel aspects of interactions between infrared photons and Dirac electrons in graphene. Commonly, drag effects are understood as friction-like momentum transfer between two coupled sub-systems. Examples include Coulomb drag between spatially separated conductors[32,33], or drag effects between electrons and phonons in a crystal[34,35]. Our data show that the notion of drag needs to be extended to the two constituents of a polaritonic quasiparticle: a superposition of IR photons and Dirac electrons. By ramping up the current in our platform, we solely perturbed the electronic constituent of the quasiparticles. The photonic component reacts by abiding to the rules of quasi-relativistic theory (Supplementary Discussion). However, the observed effect is not a mere consequence of relativistic kinematics. The plasmonic Fizeau drag in graphene is inherently a non-equilibrium and nonlinear phenomenon whose magnitude depends on the dynamics of electron-electron, electron-phonon, and electron-photon interactions of the Dirac electrons. A task for future experiments is to map the Fizeau drag for the entire SPP dispersion in order to optimize the infrared nonreciprocity for on-the-chip applications. In principle, plasmonic Fizeau drag offers means to probe the unique motional Fermi liquid effects[28] and quantum nonlocal effects[36]. Fizeau drag experiments can also be extended to double layer graphene[37] and twisted bilayer graphene[38], potentially offering intriguing opportunities to probe the physics of Fermi velocity renormalization and strong correlations[39,40] in the electronic systems. Finally, by further enhancing the drift velocity of carriers and approaching the plasmon velocity, Fizeau drag can be boosted, and plasmon instability[41,42] and amplification[43,44,45] are expected to pave the way towards realizing plasmonic emission.

## Contributions

D.A.B. and D.N.B. conceived the project. Y.D., L.X., S.Z., D.A.B. and D.N.B. designed the experiments. I.Y.P., and D.A.B. fabricated the devices. S.L. and J.H.E provided the isotopic hBN crystals. Y.D., L.X., A.S.M. and R.P. performed the experimental measurements. Y.D., L.X., R.J., F.L.R., D.A.B. and D.N.B analyzed the experimental data. Z.S., M.M.F., A.J.M. and L.S.L. developed the theoretical analysis of the experimental data with input from P.J.-H., H.G. and Z.D. Y.D., L.X., Z.S., M.M.F., D.A.B. and D.N.B. co-wrote the manuscript with input from all co-authors. D.A.B. and D.N.B. supervised the project.

## Competing interests

The authors declare no competing interests.

## Acknowledgement

Research at Columbia was supported as part of the Energy Frontier Research Center on Programmable Quantum Materials funded by the US Department of Energy (DOE), Office of Science, Basic Energy Sciences (BES), under award no. DE-SC0019443. D.A.B.

acknowledges the support from MIT Pappalardo Fellowship. M.M.F. is supported by the Office of Naval Research under grant ONR-N000014-18-1-2722. Work in the P.J.H. group was supported by AFOSR grant FA9550-16-1-0382 and the Gordon and Betty Moore Foundation's EPiQS Initiative through Grant GBMF9643 to P.J.H. The development of new nanofabrication and characterization techniques enabling this work has been supported by the US DOE Office of Science, BES, under award DE-SC0019300. This work also made use of the Materials Research Science and Engineering Center Shared Experimental Facilities supported by the National Science Foundation (NSF) (Grant No. DMR-0819762). Support from the Materials Engineering and Processing program of the National Science Foundation, award number CMMI 1538127 for hBN crystal growth is also greatly appreciated.